# Review of Explainable Decision Support and Adaptive Human–Machine Interfaces for Automation Transparency in Maritime Autonomous Surface Ships

Zhuoyue Zhang, Haitong Xu*, and Carlos Guedes Soares

***Abstract*—This paper presents a systematic review of automation transparency for Maritime Autonomous Surface Ships (MASS) across situation awareness (SA), human factors, interface design, and regulation. The analysis maps the Guidance, Navigation, and Control (GNC) architecture to shore-based operational modes and identifies concentrations of unsafe human control actions (Human-UCAs) in handover and emergency loops. Design strategies are organized across three layers: sensor-level data fusion, Human-Machine Interface (HMI) design, and engineering development processes. The review further examines diverse methods, including Human-UCA identification, quantitative measurement of trust and SA, operator workload monitoring, and the formalization of maritime regulations. Based on these findings, an adaptive transparency framework is proposed that combines real-time operator state estimation with explainable decision-support systems. The framework aims to reduce operator cognitive load and improve takeover timeliness and safety. Practical measures include intuitive risk displays, transparent model outputs with regulatory traceability, and high-fidelity Hardware/Model-in-the-Loop (HIL/MIL) simulation environments for robust operator training.**

***Index Terms*—Maritime Autonomous Surface Ships; Automation transparency; Explainable AI (XAI); Human–automation interaction; Situational awareness; Human factors.**

## I. INTRODUCTION

### *A. Maritime Automation and the Development of MASS*

Maritime transport is a major driver of international trade and economic development. More than 80% of global trade depends on maritime transport, which grew by 2.4% in 2023 to 12.3 billion tons and is projected to continue growing annually through 2029 [1]. At the same time, Artificial Intelligence (AI) is transforming transport, including maritime automation, by reshaping traditional human roles and operational practices [2], [3].

Industries are undergoing the Fourth Industrial Revolution (*Industry 4.0*), with widespread digitalization, greater connectivity, and advanced automation replacing conventional machinery. The transition to *Industry 5.0* emphasizes the *"synergy between humans and autonomous machines"*. It therefore requires attention to the *"Human-Machine Interaction (HMI) point"* and the integration of *Human-Centered Design (HCD)* principles into technological development [4], [5]. A new generation of *collaborative robots* can observe, sense, and respond dynamically to human presence, concerns, and expectations [5]. *Industrial Autonomous Mobile Robots (IAMRs)* are AI-based systems equipped with sensors that independently interpret their environment and detect obstacles without operator intervention [6].

A *Maritime Autonomous Surface Ship (MASS)* is a specialized application of IAMR technology in the maritime domain [6]. It is defined as a vessel that uses AI to operate without human supervision and improve the resilience and efficiency of maritime trade [6], [7]. Under *International Maritime Organization (IMO)* regulations, operation in international or foreign territorial waters requires flag-state approval [6]. Compared with conventional ships, MASS offers greater accessibility and safety, lower operating and crew costs, better environmental performance, more design flexibility, higher economic and supply-chain efficiency, and new military and security capabilities [8]-[10].

Autonomous vessels are being developed in research projects across Europe. EU AUTOSHIP demonstrators, including the *ZULU* pallet-shuttle barge and autonomous freight ferries tested in the Oslofjord, show varied applications of maritime automation [11], [12]. Other projects worldwide have developed autonomous ship prototypes. Yara and Kongsberg's Yara Birkeland became the world's first fully electric autonomous container ship. *ReVol*t developed a small autonomous vessel to test autonomy in short coastal shipping [8]. *MUNIN* produced findings on remote operator workload estimation and validated HMI designs within its project framework [13]. The *AAWA* project has brought together shipbuilders, equipment manufacturers, and researchers through Rolls-Royce, with the goal of deploying a fully autonomous ship by 2035 [4].

*Fig. 1 traces the IMO's shift from traditional maritime safety to MASS safety and related principles through policies and conferences. During the 20th century, the IMO built the*

This work is a contribution to the project AutoOcean - "Distributed Intelligent decision-making system of Multi Autonomous surface vehicles for sustainable ocean monitoring" (2024.12577.MIT) funded by the Portuguese Foundation for Science and Technology (Fundação para a Ciência e a Tecnologia-FCT). This work also contributes to the Strategic Research Plan of the Centre for Marine Technology and Ocean Engineering, financed by FCT under contract UID/00134/2025 (https://doi.org/10.54499/UID/00134/2025).

Zhuoyue Zhang is with the Civil and Architectural Engineering Department, KTH Royal Institute of Technology, Stockholm, 11428 Sweden (e-mail: zhuoyuez@kth.se).

Haitong Xu and Carlos Guedes Soares are with Centre for Marine Technology and Ocean Engineering (CENTEC), Instituto Superior Técnico, Universidade de Lisboa, Lisbon, Portugal (e-mail: haitong.xu@centec.tecnico.ulisboa.pt).

*(Corresponding authors: Haitong Xu)*

*main regulatory framework for seafarer safety, collision prevention, and marine pollution. Key conventions were the International Convention for the Safety of Life at Sea (SOLAS) [15], the International Convention on Standards of Training, Certification and Watchkeeping for Seafarers (STCW) [16], and the Convention on the International Regulations for Preventing Collisions at Sea (COLREGs) [17].* With Industry 4.0, MASS *bec*ame a critical technology and a *"*prime growth engine*" fo*r the transition to Industry 5.0 [4], [18]. *At* its 98th session *in June 2017*, the Maritime Safety Committee (MSC) launched the *first* Regulatory Scoping Exercise (RSE) for MASS. MASS *then* gai*ned* bro*ad* attention and bec*a*me a main item *at late*r MSC *sessions.*

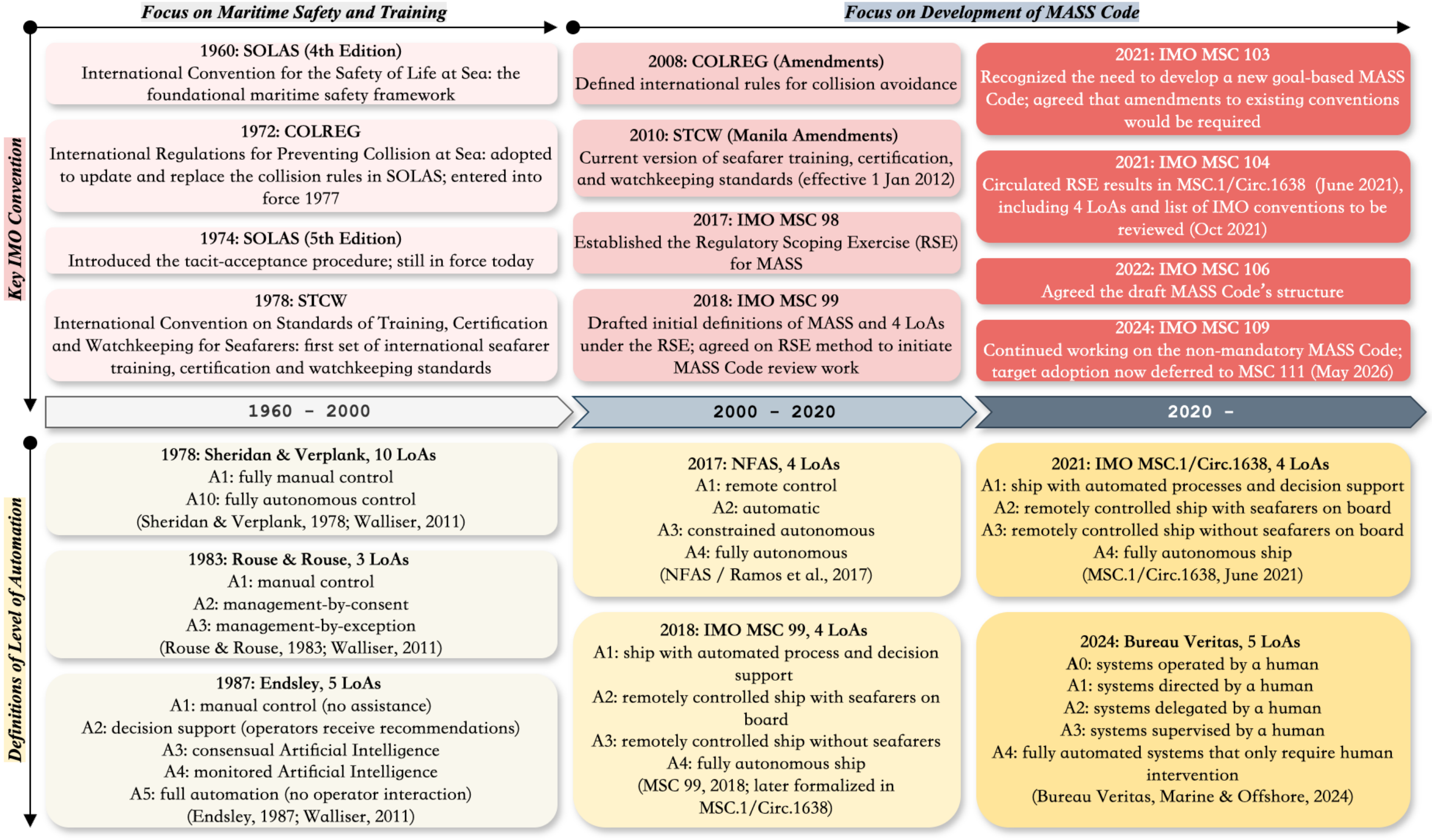

**Fig. 1.** Evolution of IMO focus and the MASS Code timeline [14]-[17].

At its 99th session, the MSC drafted definitions of MASS and *Levels of Autonomy (LoA)*. It also agreed on the RSE methodology to begin reviewing the *MASS Code*. At its 103rd session, the MSC identified the need for a new non-mandatory MASS Code. Drafting continued at the 109th session in December 2025. Finalization and adoption remain scheduled for MSC 111 in May 2026 [18]-[21].

## *B. AI, Autonomous System, and LoA*

The maritime domain is changing rapidly through advances in automation, robotics, multisensor perception, AI, and the convergence of digitalization and connectivity [22]. AI is central to autonomy because it supports intelligent perception, decision making, and control in MASS.

The MASS autonomous navigation system has two main components: a decision-making system and an action-taking system. Onboard computers use AI algorithms, including machine learning and deep learning, to process sensor inputs in real time. This supports situational awareness, collision avoidance, and the emulation of human behavior. Together with communication networks for remote operation, these capabilities enable efficient autonomous navigation with limited human intervention [8].

*Level of Automation (LoA)* is fundamental to autonomous systems because it defines how responsibility is divided between humans and automation. Sheridan and Verplank introduced LoA in 1978 as a 10-level framework. It ranges from fully manual tasks with minimal automation at Level 1 to fully computer-controlled tasks with maximum automation at Level 10 [23]. Although LoA criteria have since evolved (Fig. 1), the core principle remains unchanged. Manual control is the lowest level and full automation is the highest. For autonomous ships, the Norwegian Forum for Autonomous Ships (NFAS) proposed the first maritime LoA classification in 2017. Its four-level scheme was designed for autonomous merchant vessels [24]. The IMO then formally defined four LoAs specific to MASS at the 99th MSC session in 2018. The definitions were refined and officially adopted through MSC.1/Circ.1638 in June 2021. A1 describes a ship with automated processes and decision support. A2 describes a remotely controlled ship with seafarers on board. A3 describes a remotely controlled ship without seafarers on board. A4 describes a fully autonomous ship. This standardized classification is now widely used in MASS regulatory development and discussions and is the main reference for ongoing MASS Code development [19], [20], [129].

### C. Contributions of this review

This review advances MASS research in four complementary areas. Its novelty lies in linking Human-UCAs, trust and situation awareness, HMI transparency, validation maturity, and governance in a unified sequence connecting problem identification, design, and validation, rather than treating MASS transparency as a general descriptive theme.

First, it establishes a detailed operational map linking the Guidance, Navigation, and Control stack to shore-based Remote Supervision and Remote Control. The map pinpoints Human-UCA hotspots, especially in handover and emergency loops, and integrates STPA-Cog with IDAC to derive safety constraints.

Second, it consolidates an evidence based measurement toolkit combining validated trust and situation awareness instruments, behavioral and performance indicators, EEG and fNIRS measures, and preregistered analyses across high fidelity simulations and testbeds using hardware, software, or humans in the loop, as well as digital twins.

Third, it proposes an actionable layered framework that organizes transparency strategies across SA acquisition, HMI presentation, and engineering processes. It translates principles into CPA/TCPA risk widgets centered on charts, robustness heatmaps, displays tracing COLREG rules, resilient interaction design, and standardized data sharing via VDES.

Finally, it introduces an adaptive transparency framework linked to governance. By combining estimates of operator state with explainable decision support, the framework regulates what to explain and when to reduce cognitive overload and improve takeover timeliness. It also connects technical solutions to transparency in safety cases, the emerging MASS Code, and standardized route and intent communication.

## II. Transparency As A Human-Centric Approach In MASS

### A. Working shift from Seafarers to RO

Seafarers traditionally use sailing skills and maritime experience to operate ships and support global trade [7]. MASS creates a major shift by combining human intelligence with advanced machine capabilities. This shift redefines maritime roles by combining traditional skills with technological expertise. The result is a hybrid profession that is forward-looking while grounded in established maritime knowledge [25].

Although MASS can operate at different LoAs, human intervention remains necessary. Even at Level 4, defined by IMO MSC.1/Circ.1638 as a fully autonomous ship without crew on board, humans still support control and monitoring. These tasks are usually performed remotely from shore-based facilities such as the Shore Control Center (SCC) [6], [7], [9], [127]. Data collection, prediction and processing, decision making, and action execution are increasingly moving to remote environments. Human operators remain essential backups when automated systems fail [7], [26]. This shift is creating a new generation of seafarers, often called Seafarer 4.0, who need digital skills to manage and supervise autonomous vessels effectively [4].

In MASS operations, this emerging role is more precisely termed the Remote Operator (RO). The 106th MSC defines an RO as "any person, including the Master, with recognised or certifiable experience who is engaged in the remote operation of a Remotely Operated Unmanned Vessel (ROUV)" [20], [108]. ROs usually work in a Remote Operation Center (ROC), defined as "either a shore-based location which is permanent or mobile, or a manned vessel from which a Remotely Operated Unmanned Vessel is operated" [20]. Their responsibilities depend on vessel autonomy. For high-LoA MASS, ROs mainly supervise remotely. At lower LoAs, they actively control vessel functions. When the autonomous system cannot resolve an unexpected situation, the RO intervenes remotely to return the vessel to a safe state [24].

### B. Importance of Human-centric Design: Human error, XAI, Trustworthy

Automation works best when systems are predictable and place limited cognitive demands on operators. Excessive complexity can weaken human–automation interaction and operational effectiveness [27]. Users should therefore be involved throughout system design, especially for ROCs. Performance in these settings depends on cognitive load, situational awareness, training, skill levels, and teamwork. Human reliability in emergencies remains a major concern and requires further research and interventions at system level [28].

*Human-Centered Design (HCD)* addresses these challenges by placing user needs at the center of system development. Unlike technology-driven design, it adapts systems to human capabilities rather than requiring people to adapt to technology. ISO 9241-210 defines HCD as "*an approach to systems design and development that aims to make interactive systems more usable by focusing on the use of the system and applying human factors, ergonomics and usability knowledge and techniques*" [27]. For MASS, HCD is essential for aligning system functions with operator skills and safety requirements.

HCD is also central to the developing MASS Code, which treats human factors as a core concern. The current draft has three sections: *Definitions*, *Principles*, and *Objectives*. Part 2 includes a provision on the "*Function and role of the operational environment and human factors*". It covers operational context, risk assessment, system design, connectivity, and human elements [18]. Early integration of human factors helps engineers build usable and resilient systems [29]. A human-centric perspective also supports analytical frameworks for interactions among *Risk-Influencing Factors (RIFs)*, providing insights for crew training and vessel design [30].

Trust is critical to human–machine interaction and rests on "*persistence, competency, and fiduciary responsibility*" [23]. Although AI and ML are central to MASS, complex models such as neural networks can undermine operator trust. Their interconnected parameters, lack of interpretability, and potential biases from incomplete or skewed training data make decisions opaque. This "*black box*" makes it difficult for ROs and nearby seafarers to understand or predict system actions [31]. Discrepancies between human and system situational awareness have also been shown to increase cognitive stress and impair decision-making [32]. Moreover, applying COLREGs follows the human-like sequence of "*assessing the situation—deciding on the applicable rule—taking action*", leaving RO decisions vulnerable to subjective error. Human error has long been recognized as the leading cause of maritime accidents and is estimated to account for 89%–96% of collisions [8], [24], yet human-related risks have historically received insufficient attention in system design.

Human-centric AI design must prioritize *explainability* to reduce these risks. Operators need to know what the system is doing at any time [27]. This requirement underpins *Explainable Artificial Intelligence (XAI)*, also termed "*automation transparency*" in autonomy and robotics [8], [31]. Reference [34] identified the "*Glass Box*" as a key research area. AI systems should present decisions, their reasoning, and alternative actions [35]. This approach, called *AI Decision Transparency*, concerns how autonomous systems communicate internal logic when making decisions or supporting them.

Greater transparency can prevent misunderstandings and reduce automation-induced incidents by informing operators about current and future system actions [27], [36]. Transparency also shapes trust by helping users understand internal system processes. If users cannot understand or explain automated decisions, accountability becomes difficult, especially when operators remain responsible in critical situations [36]. Improving transparency therefore supports both trust calibration and clear responsibility after incidents.

## III. Research Design

### A. Data Collection and Screening

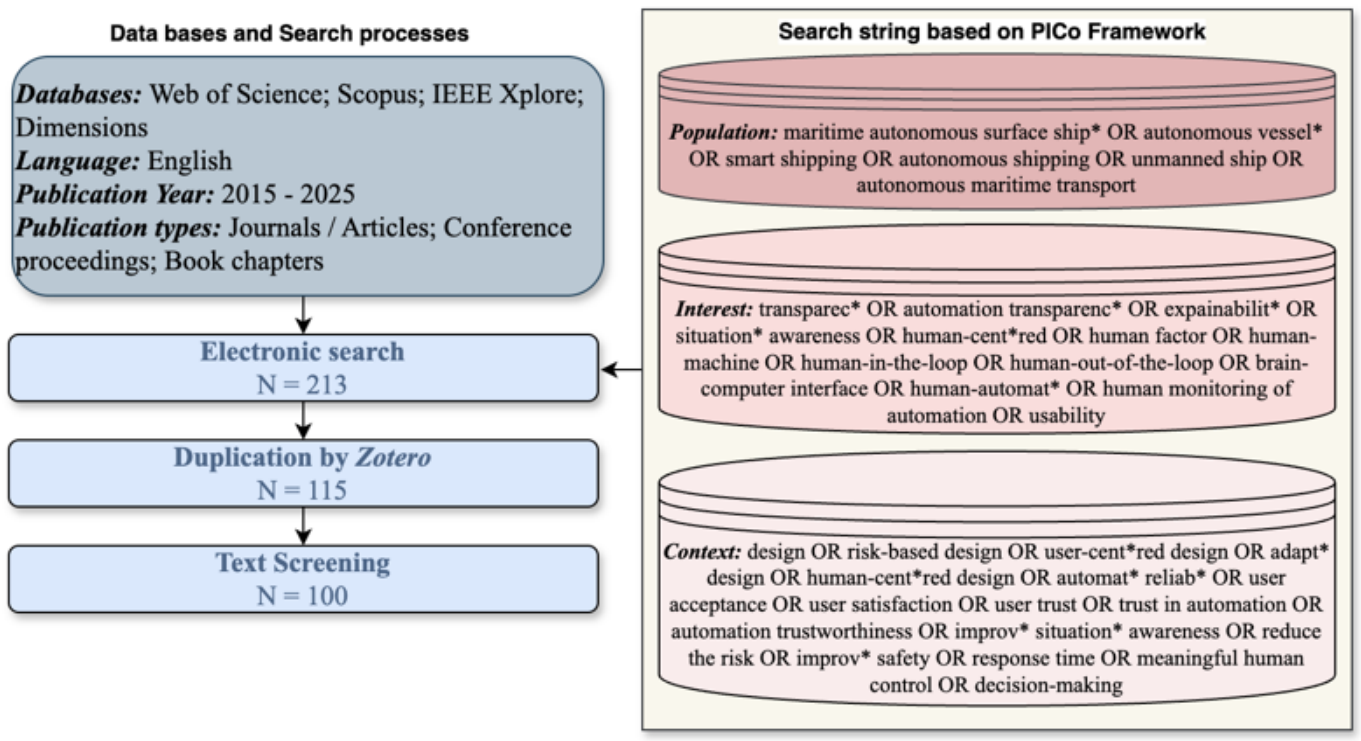

**Fig. 2.** Screening pipeline and PICo-based query.

A PRISMA-style identification and screening process ensured review reproducibility. Web of Science, Scopus, IEEE Xplore, and Dimensions were searched for English-language journal articles, conference papers, and book chapters published from 2015 to March 2025. The PICo-based query [37] was: ("maritime autonomous surface ship*" OR "autonomous vessel*" OR "unmanned ship*" OR "autonomous shipping") AND ("automation transparen*" OR explainab* OR "situation awareness" OR HMI OR "human-machine" OR "human-in-the-loop" OR "out-of-the-loop" OR "trust calibration" OR "remote operator").

Inclusion required direct relevance to MASS or autonomous or remotely operated vessels and at least one of ROC/RO work, transparency, explainability, SA, trust, HMI, safety validation, or governance. Exclusions covered non-English records, publication outside 2015 to March 2025, non-peer-reviewed or non-authoritative sources, generic non-maritime AI/robotics, and no substantive human–autonomy discussion. The search found 213 records. Removing 98 Zotero duplicates left 115, and screening removed 15 out-of-scope records, leaving 100 studies.

Although automation transparency has tens of thousands of publications, limiting the scope to MASS reduces the number to about one-fiftieth. This shows both a major research gap in MASS transparency and the strong coverage and representativeness of the 100 studies reviewed.

### B. Descriptive Analysis

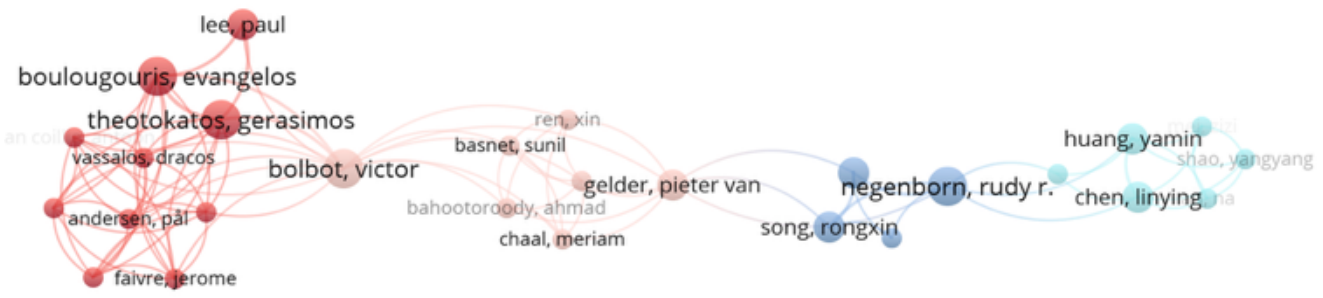

**Fig. 3.** Co-authorship network of authors (VOSviewer).

Fig. 3 maps three closely connected co-authorship communities linked by a few bridging scholars: naval architecture and safety (Theotokatos, Bolbot, and Boulougouris), control and optimization (Negenborn and van Gelder), and data-driven navigation and perception (Huang and Chen).

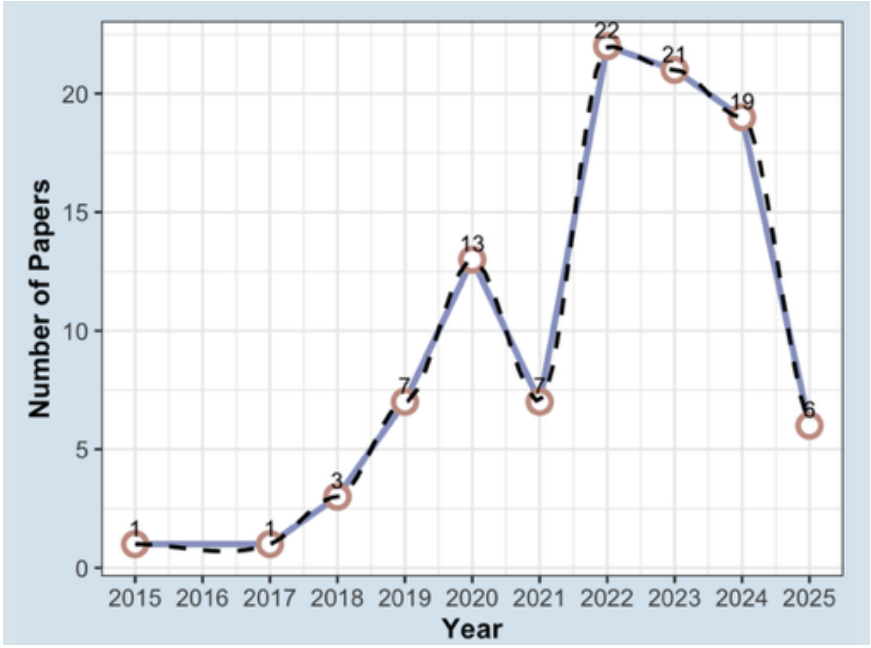

**Fig. 4.** The growth trend of published paper numbers in decade years.

Fig. 4 shows annual publication counts from 2015 to 2025, with 2025 data recorded through March. The reviewed studies increased gradually in the early years and more sharply after 2020. This trend reflects growing academic interest in MASS transparency alongside the maturation of autonomous navigation technologies and progress in IMO regulatory discussions.

Fig. 5 presents the VOSviewer keyword co-occurrence network, which identifies several thematic clusters. The largest centres on "*situation awareness*" and links strongly to "human factors", "automation", and "safety". Another connects "automation transparency" with "trust" and "human–automation interaction", while smaller clusters cover "collision avoidance" and "risk assessment", reflecting a focus on safety assurance. Together, these patterns show that MASS transparency research seeks to expose autonomous system states and strengthen trust and decision support through improved information presentation and interface design, thereby reducing navigational risk and supporting collaboration between humans and systems in increasingly autonomous maritime operations.

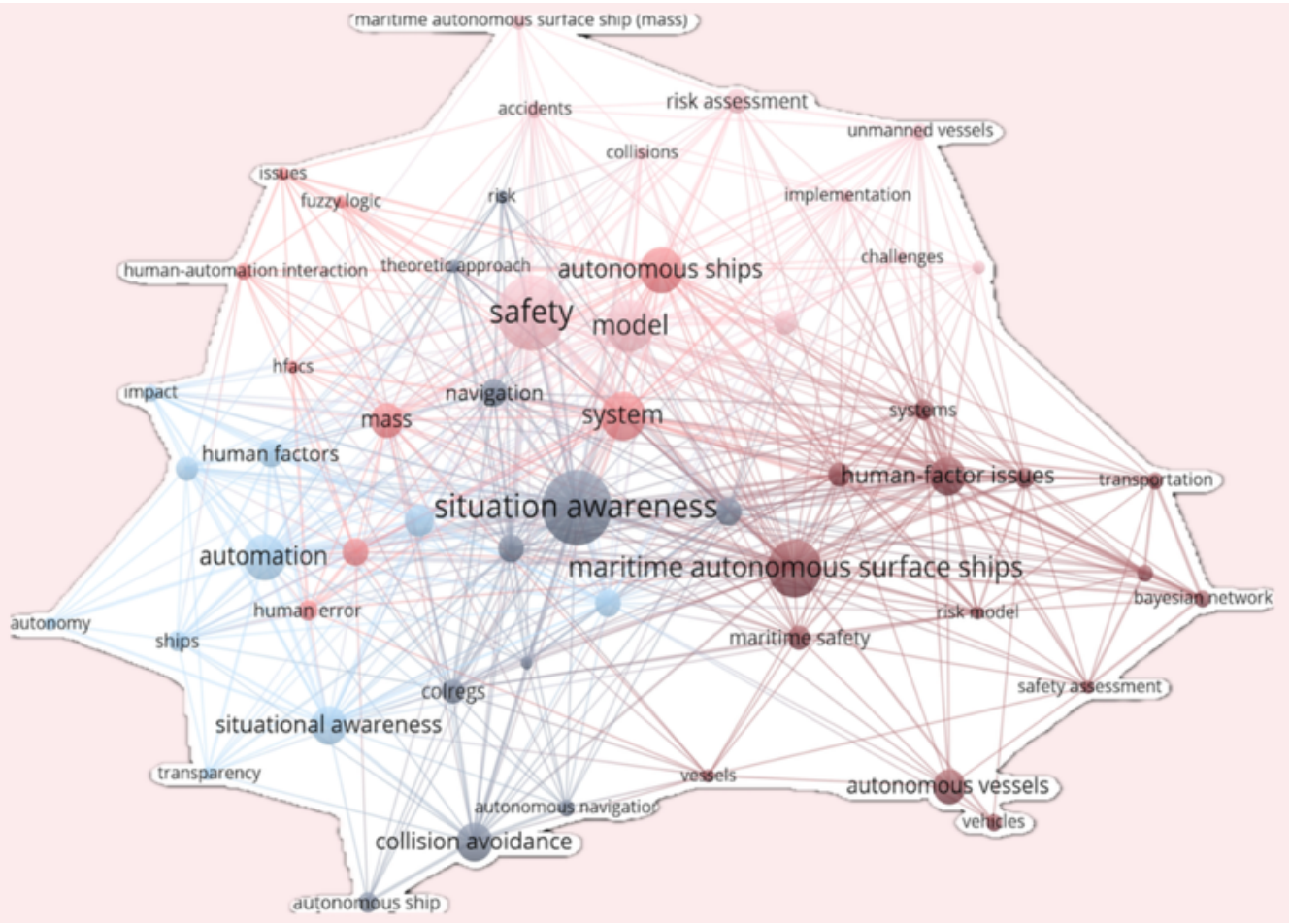

**Fig. 5.** Keyword co-occurrence network (VOSviewer).

## IV. Results And Discussion

The results are organized to follow the review logic from problem identification to design strategy, validation maturity, and governance. This structure first identifies where transparency failures arise in MASS operations, then synthesizes design responses and the evidence needed to validate them.

### *A. Transparency in MASS operation*

#### 1) General architecture and operational process of MASS

The general architecture of MASS consists of a *Guidance, Navigation*, and *Control (GNC)* system [38]. Fig. 6 shows the overall system architecture, operational workflow, and the corresponding process of human error identification.

##### a) Navigational layer: Sensors, Sensor fusion, Situational Awareness

In the navigational layer of MASS, *situational awareness* (SA) is one of the most critical functions. SA is formally defined as "*the perception of the elements in the environment within a volume of time and space, the comprehension of their meaning, and the projection of their status in the near future*" [27]. In conventional shipping and autonomous maritime systems, SA is vital for identifying, monitoring, and anticipating potential threats and environmental conditions within a defined spatial and temporal scope.

Traditional vessels mainly use onboard sensors to collect data about the environment and the ship. In contrast, MASS require a wider and more redundant sensor network to support SA. These sensors provide critical data about vessel state, such as position, speed, and heading. They also provide data about environmental conditions, such as wind, currents, and waves, and surrounding traffic, such as the trajectories and classifications of nearby ships. Common sensors include *day and night vision cameras* for high-resolution environmental imaging. *LiDAR and LADAR systems* powered by artificial neural networks support object recognition. Other sensors include *sound detection sensors, weather instruments,* and *underwater sensors.* Sensor fusion integrates and analyses these different data streams in real time.

*AUTOSHIP* proposes adding shore-based infrastructure to sensor networks to improve cost efficiency [11]. Robust communication technologies, including *GNSS*, *IMU, AIS,* and *ECDIS*, allow data to support several MASS and ROCs at once. Beyond physical sensors, fully autonomous MASS increasingly use extended SA tools in ROCs, including VR and AR. Immersive 3D displays allow ROs to explore the vessel's surroundings and detect dynamic and static objects. Compact UAVs with vertical take-off capability can also scan ahead of the vessel and extend the operator's view. These HMI and SA technologies enrich real-time awareness and support RO decisions [8]. The navigational layer also includes smart ship infrastructure, remote control, and autonomous operation, which are expected to transform maritime transport [22].

##### b) Guidance & Control Layer: RCM, RSM, Operational Loop

In the navigational layer, SA converts fused inputs, including vessel state, environmental maps, and dynamic object data, into high-level cognitive outputs. These outputs answer *"Where am I?"* and *"What is happening around me?"* and support RO decisions. Reference [39] divides ROs into the *Operational Group* and the *Navigational Group*. These groups correspond to two operating modes. In *Remote Control Mode (RCM),* the RO controls the vessel in real time. In *Remote Supervision Mode (RSM)*, the RO monitors the vessel and intervenes only when needed [38]. ROs perform four core tasks: *Monitoring, Supervision, Intervention,* and *Direct Control*. These tasks form an action-making loop [39], as follows:

*i) Monitor and Alarm: ROs continuously monitor system status and respond to any events that trigger high-priority alerts.*

Under RSM, two types of RO roles exist. (1) A backup RO evaluates incoming navigational information and decides whether to take over control based on system status. (2) A passive-control RO remains in standby until the system explicitly requests human intervention. Once either RO type assumes control, the operational mode shifts from RSM to RCM. If no intervention is deemed necessary, the system remains in RCM [38].

*ii) Wait for Auto-Recovery: the system initiates an auto-recovery process within a predefined time window and attempts to resolve the situation by generating a solution.*

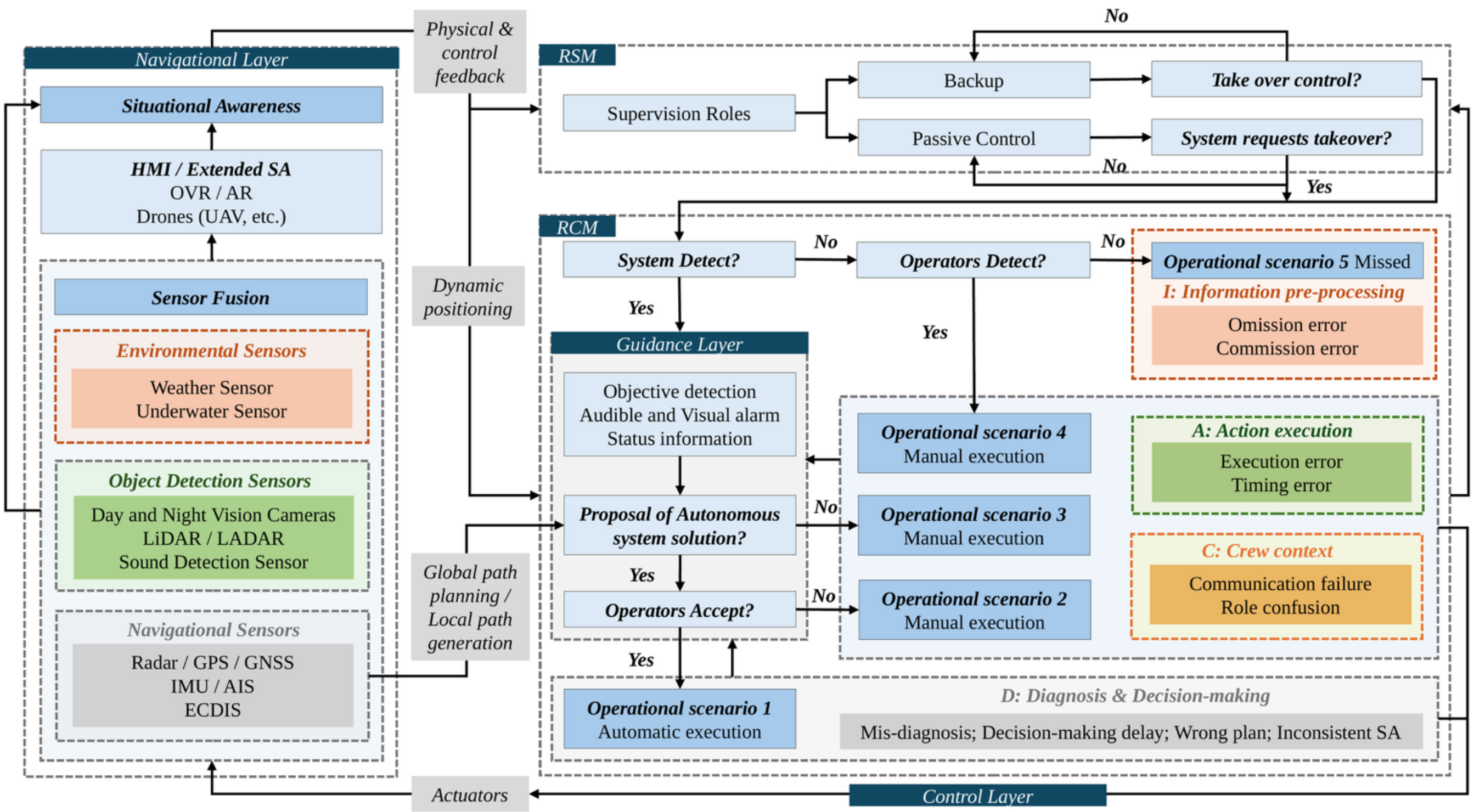

**Fig. 6.** Architecture and Operational scenarios of MASS (adapted from [8])

When the operational mode changes from RCM to RSM, the guidance layer uses environmental and vessel state data from sensors for global path planning and local path generation [8]. The system evaluates several candidate routes or maneuvering strategies, presents explainable decision outputs, and recommends an autonomous solution.

*iii) Analyze and Decide: If no valid solution is provided, the RO formulates an action plan. If a solution is presented, the RO evaluates its reliability and decides whether to accept or override it.*

There are five operational scenarios during this period [24], as presented in Table I.

TABLE I

THE SUMMARY OF FIVE OPERATIONAL SCENARIOS

| | |
|---|---|
| *Operational scenario 1* | Automatic execution with successful system detection and approved proposal of autonomous system solution |
| *Operational scenario 2* | Manual execution with successful system detection and rejected proposal of autonomous system solution |
| *Operational scenario 3* | Manual execution with successful system detection and no proposal of autonomous system solution |
| *Operational scenario 4* | Manual execution with operator detects only |
| *Operational scenario 5* | Missed / Latent hazard |

*iv) Trigger Action: the RO consults the group leader and executes the selected plan.*

Once a decision is confirmed at key operational nodes, commands for the five defined operational scenarios are compiled and sent to the control layer.

The control layer then sends specific rudder or thrust instructions to the actuators for execution [38].

*v) Loop: If the issue remains unresolved, the system repeats these steps until it is resolved.*

The control layer sends execution feedback to the navigation and guidance layers. Updates to SA and dynamic positioning data form a feedback loop that continuously adjusts decisions and actions. After the task is completed, the system returns from RCM to RSM.

2) **Human Error Identification**

**a) Human-UCAs Identification Methods**

*Unsafe Control Actions (UCAs)* are control commands that may cause hazardous outcomes in specific operational contexts and worst-case environmental conditions. Human error is treated as a UCA when the operator's action does not match the expected system state, procedural norms, or the behavior of other crew members. These mismatches are called *Human-UCAs.* The same action may be correct in one situation but wrong in another. Human-UCAs must therefore be identified using defined *context variables (CVs)* [24], [38].

Across the 100 studies, human-error identification commonly combines system analysis methods, including *STPA, STAMP, and HFACS*, with cognitive models. Together they classify and interpret collision and *control scenarios (CSs)*. Fig. 7 shows the integrated framework.

*i) Cognitive models*

The *IDAC operator cognitive flow model* is most often combined with system analysis methods. References [24] and [38] state that "*IDAC is composed of models of information pre-processing (I), diagnosis and decision-making (D), action execution (A), and a crew context (C)*". Other cognitive models are also used.

Reference [36], for example, applied the *Perceptual Cycle Model (PCM)* to human–system errors because it places the decision maker's cognitive process within the system context.

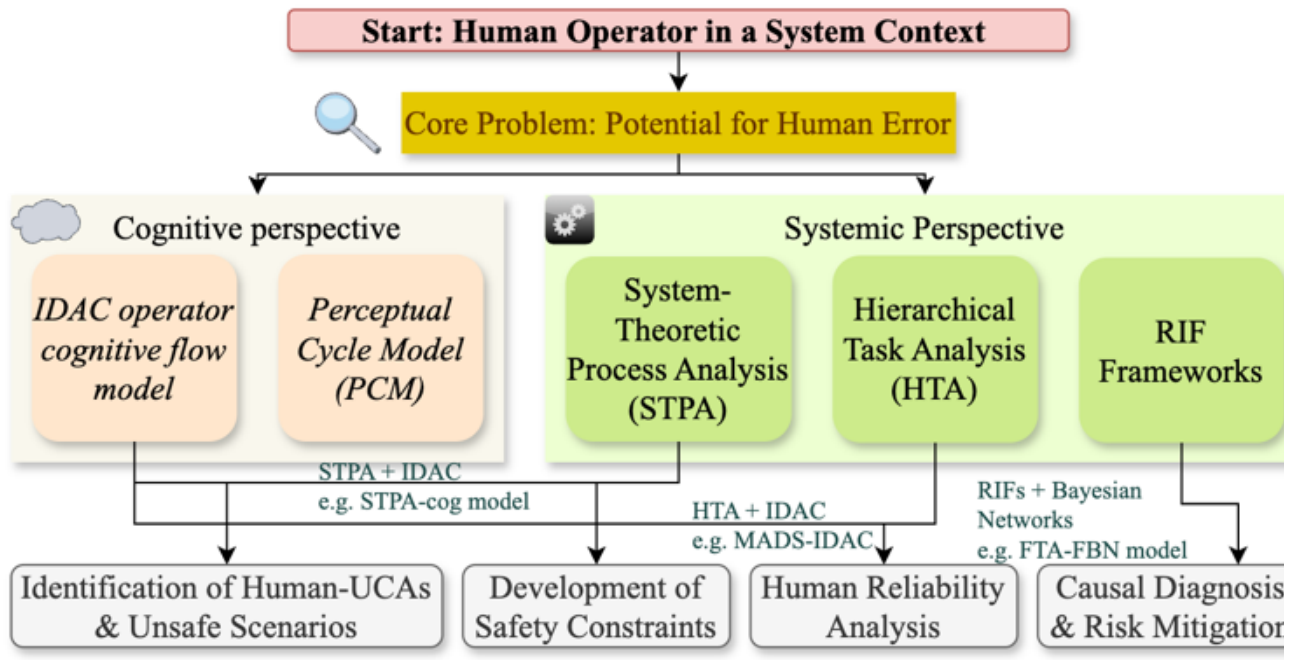

**Fig. 7.** Integrative framework for human-error analysis in MASS

*ii) System and task analysis approaches*

System and task analysis methods help identify human errors. Reference [38] introduced *STPA-cog,* which combines *System-Theoretic Process Analysis (STPA)* with the IDAC cognitive model. It treats accidents as outcomes of UCAs in complex systems and examines control and feedback loops. Standard guidewords identify actions that are *not provided when required*, *provided when not required*, *provided too early* or *too late*, *stopped too soon* or *applied too long*, or use the *wrong type or value*. These guidewords generate CSs and safety constraints for each control action. STPA is recognized as one of the most promising tools for MASS hazard analysis [40] because it does not require large historical safety datasets. *Bayesian Network*s *(BN*s*)* share this advantage [41], [42]. Reference [43] increased analytical specificity by integrating STPA with *Operational Design Domain (ODD)* definitions. The resulting method identifies systemic risks more rigorously by defining the environmental and operational conditions in which the autonomous system is designed to operate.

*Hierarchical Task Analysis (HTA)* decomposes complex operations into hierarchies of goals and subgoals [24]. Combined with IDAC, it maps each lower-level task to the I/D/A phases to identify *human failure events (HFEs)*. H*uman reliability analysis (HRA)* then follows four steps: task decomposition using HTA, cognitive mapping, error identification from mismatches, and quantification with models such as THERP or SPAR-H [24]. Simulation extensions, including the *MADS-IDAC integration* proposed in [44], further support error prevention at the crew level.

Customized *Risk Influencing Factor (RIF)* frameworks extend these methods. Reference [45] identified 23 human-related RIFs across six categories: cognitive, psychological, error types, situational awareness, experience and training, and cooperation. Expert judgement mapped them to four voyage phases: voyage planning (V), berthing and unberthing (B), port approach and departure (P), and open-sea navigation (O). For high-risk berthing and unberthing, [46] recommended combining static and dynamic data to capture temporal risk correlations. Bayesian Networks are widely used to model RIF interdependencies and support causal inference for specific accident types [30]. Reference [47] proposed a hybrid *Fault Tree Analysis–Fuzzy Bayesian Network (FTA-FBN)* model. FTA first establishes causal relations, which are then mapped into a Bayesian Network for fault diagnosis and identification of key contributors to MASS collision risk under uncertainty. References [48] and [49] further advocated a global perspective in data-driven collision risk analysis to explain accident causality beyond localized scenarios.

### b) Human-UCAs in RSM & RCM

Human-UCAs mainly arise in two parts of MASS operations. The first is deciding whether to take control during the transition from RSM to RCM. The second is manual risk across the five RCM scenarios. During decision making, RCM contains about 51% more CSs than RSM, showing that diagnosis and decisions impose greater cognitive demands in RCM [38].

Under RSM, the RO only decides whether to switch to RCM [38]. The *mode-switch command* is therefore the main control action for applying STPA guidewords. Human-UCAs include unnecessary takeover, failure to take over when required, and takeover at the wrong time.

Fig. 7 maps potential human errors in RCM scenarios to the four IDAC cognitive blocks. Critical points include "*operators detect: scenario 5*", "*operators accept: scenario 1*", and manual execution in scenarios 2, 3, and 4.

- *i)* In "*operators detect: scenario 5*", both the operator and system may miss the event because information is incorrect, late, or unreadable. This is an I (Information pre-processing) failure that can cause omission or commission errors.
- *ii)* In "*operators accept: scenario 1*", the operator may exhibit feedback delay or confirmation bias caused by inadequate decision planning or incomplete/inaccurate situational awareness, resulting in either a failure to identify abnormal automation behavior (*miss*) or an incorrect acceptance of an inappropriate system plan (*false alarm*) [23]. These errors map to the D (Diagnosis and Decision-making) block.
- *iii)* Manual execution in *scenarios 2, 3, and 4* can involve incorrect rudder angle or thrust, timing errors, communication failures, or role confusion when actions are triggered. These errors relate to the A (Action execution) and C (Crew context) blocks [38].

### c) Contributing Factors to Human-UCAs

Using STPA guidewords, [38] identified Human-UCA categories and immediate causes such as poor decision planning and incomplete SA. These immediate causes often reflect two deeper problems. System-level factors limit observability through poor information acquisition and presentation. Human-level factors hinder performance through ineffective interaction with automation and misaligned trust calibration.

*i) System-level factors: SA information collection and display*

DSS information quality and presentation directly affect how operators perceive, understand, and respond to navigation situations [36]. Sensor drift, noise, and poor calibration can reduce SA data reliability and prevent both the system and RO from making accurate decisions in real time [8].

Interface visualization can introduce complications in ROCs. Large displays provide live onboard video and a view from the vessel, but only approximate the actual bridge view. Transmission latency and degraded image quality can impair situational interpretation and lead ROs to make different decisions in identical ship encounters. Reference [50] demonstrated this in a UiT bridge simulator by comparing ARPA radar information with ARPA data supplemented by a camera with a limited field of view. In the second condition, the operator chose a more conservative but safer evasive maneuver, reducing speed before sharply altering course. The operator reported that a full bridge view would have enabled earlier and more effective assessment [50]. The study shows that variations in SA conditions directly affect RO decisions and subsequent actions.

*ii) Human-level factors: Trust, Cognition, and Decision-making*

A) *Clumsy automation risk*

Emergency handover failures are common in highly automated systems because passive monitoring can leave operators "out of the loop" [36]. The same risk arises in MASS, where ROs need sufficient time and contextual information to assess emergencies, intervene, and execute evasive actions to avoid collisions. Table I shows that scenarios 2, 3, and 4 require ROC intervention. During LoA transitions, the speed and volume of sensor data may exceed RO processing capacity, creating clumsy automation and a cognitive bottleneck that coupling risk can intensify [51]. In opaque unsafe scenarios, limited transparency about system status and decision rationale can fracture trust and increase cognitive load. Operators may also misunderstand automation limits and miss the narrow window for retaking control [36]. The resulting human errors include execution errors, timing errors, and communication failures [27], [29].

B) *Poor trust calibration*

Trust calibration aligns an operator's trust in automation with trust in manual control and is essential to reducing human error. Signal Detection Theory (SDT) evaluates calibration by classifying responses as hits, misses, false alarms, or correct rejections [23]. Well-calibrated operators appropriately rely on or reject automation, producing more hits and correct rejections. Poor calibration produces more misses and false alarms, leading to automation overuse or disuse.

C) *Lack of ship sense*

Remote control can reduce realism when interfaces resemble game environments, as shown in full bridge cyberattack simulations [52]. Away from the vessel, ROs lose "ship sense", including motion cues and external visibility. Prolonged monitoring can also increase boredom, while handover difficulties may cause accidents in uncrewed operations [36].

D) *Cognitive overload*

Large volumes of onboard sensor data increase RO workload [36]. Under the "*six-ships concept*", one RO could supervise up to six vessels in open-sea areas with little traffic, spending about 10 minutes per vessel to ensure hourly inspection [53]. Yet current HMIs struggle to support effective monitoring of even one uncrewed vessel. Supervising several ships may therefore exceed cognitive capacity, causing overload and poorer decisions [36].

3) **How to reduce human-UCAs: From transparency, to trust, to explainability**

How to better address Human-UCAs remains a key challenge for human-centered design in MASS. As noted in Section 2.2, improving system transparency and explainability is crucial for enhancing operator trust and reducing human error, which can be achieved through the implementation of *Explainable AI (XAI)*.

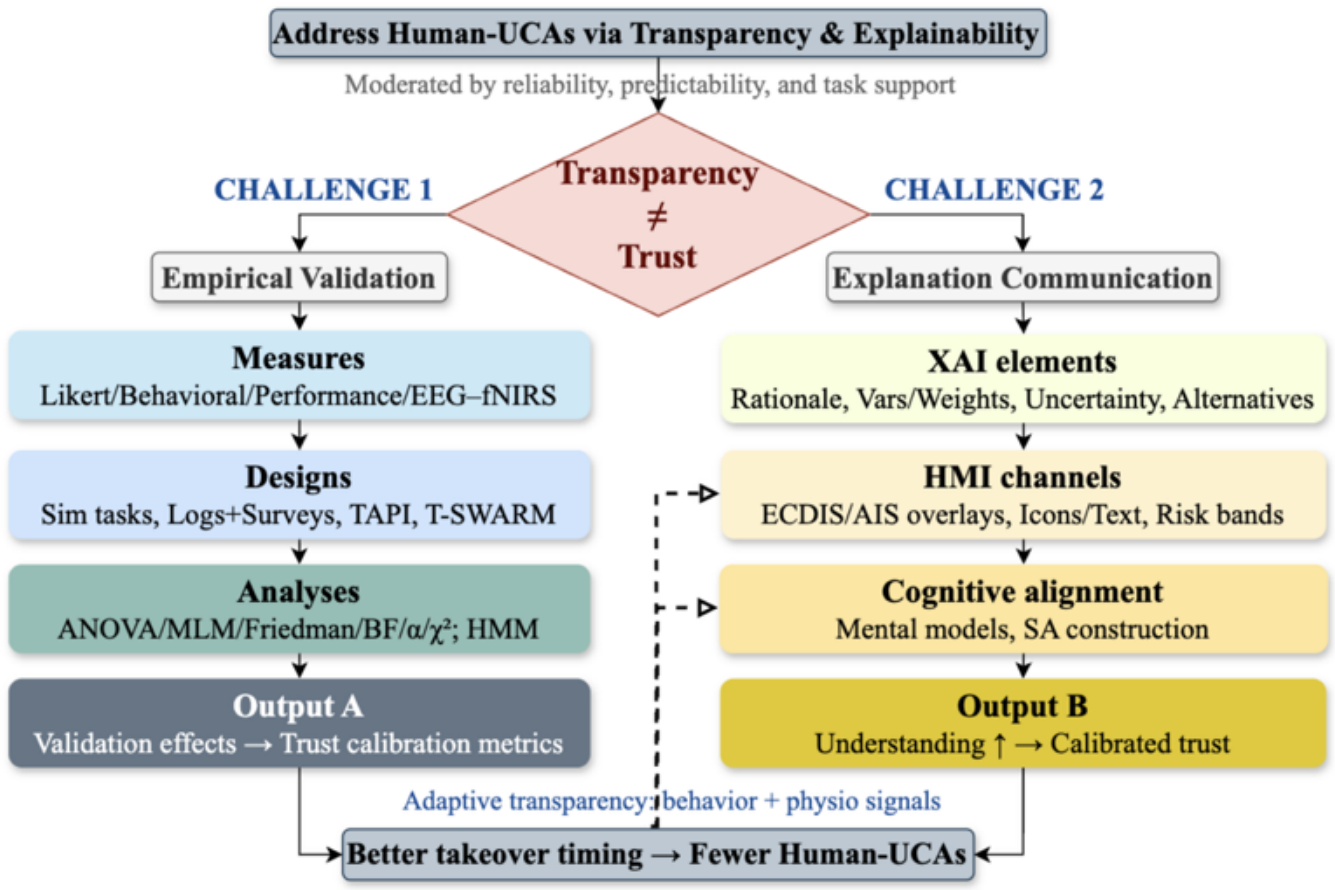

**Fig. 8.** From transparency to trust calibration in MASS: dual-challenge framework

XAI aims to make AI systems understandable by aligning explanations with users' mental models, thereby supporting trust and effective interaction [33]. *I*n a simulation using ECDIS and AIS, [35] showed that decision support systems should communicate recommendations transparently, make alternative actions easy to execute, and explain how SA is constructed. These findings underscore the need for further research on AI decision transparency strategies. Greater understanding of system decision processes also improves trust calibration [54]. Participants in [36] specifically identified the variables and weightings behind recommendations as critical to trust formation, confirming that transparency directly shapes trust.

Transparency, trust, and explainability do not have a simple or mutually reinforcing relationship. Explanations improve operator understanding more consistently than trust. Their effect on trust depends on reliability, predictability, task support, and prior training biases [31], [36], [55]. Greater transparency therefore does not always produce calibrated trust. Fig. 8 summarizes two main research challenges:

(1) ***Non-empirical validation of trust****: Reference [31] noted that most guidelines for improving trust and transparency rely on subjective experience rather than empirical evidence, resulting in what they called "unvalidated guidance".*

(2) ***Lossy communication of explanations****: Reference [33] emphasised that even a fully transparent AI is not necessarily explainable. Explanations must be perceived and accepted by users, and failure in this step can remain a bottleneck to building trust in autonomy.*

For challenge (1), studies use quantitative and qualitative methods to test whether greater MASS transparency increases RO trust. Trust in automation depends on predictability, dependability, and trustworthiness [36], while reliable questionnaires should use established human factors instruments that include trust measures [31]. Most studies quantify trust with Likert questionnaires and statistical analysis. Reference [56] used a situation awareness questionnaire to examine subjective ratings, decisions across scenarios, and the timing of decision handling. Reference [57] combined operational logs recording when, why, and how operators switched to manual control with interviews after the task and Likert evaluations to study the timing from monitoring to takeover and associated trust feedback. Reference [58] analyzed shore control experiments to quantify human errors during collaboration with automation, providing empirical validation of theoretical error classifications. Qualitative methods examine operator expectations and cognition. Reference [29] used Think Aloud Protocol Interviews (TAPI) to study expectations of transparency and HMI design, while [59] used the Trust Schema World Action Research Method (T-SWARM) to examine how trust shapes RO decision processes.

Statistical methods include *Analysis of Variance (ANOVA), Mixed Linear Models,* the *Friedman Test, Bayes Factor, Cronbach's Alpha,* and *Pearson's chi-square test* [31]. Simulation studies also combine sensitivity analysis with scenario-based risk prediction and diagnosis. For example, [60] used a *Hidden Markov Model (HMM)* to predict navigation states and evaluate reliability in cooperative navigation between humans and machines.

Physiological and psychological metrics also validate operator trust and cognitive load. These include *electroencephalogram (EEG)* workload assessments and *functional near-infrared spectroscopy (fNIRS)* studies of mental workload [61]. Evidence that trust differs between individuals can inform adaptive transparency settings and help maintain an appropriate level of operator trust [36].

For challenge (2), *how to efficiently communicate explanations while improving MASS transparency and operator trust*, studies show that explanations do not always significantly increase trust but consistently improve *understanding* [31]. In human–robot collaboration, understanding human cognition and decision making is essential to transparent and explainable HAI design [6], [26]. H*uman-centered design (HCD)*, particularly support for SA, can reduce new human errors [27]. The HMI therefore plays a central role in communicating explanations and should reflect users' mental models of system decisions [33]. Section 4.2.1 presents specific HMI design methods.

Before discussing specific HMI design methods, these findings can be organized into an adaptive transparency framework. Building on the mixed effects of explanations on trust, the proposed adaptive transparency framework regulates explanation detail, timing, and modality according to operational risk and RO state. Its input layer combines system status (GNC mode, sensor confidence, COLREG rule trace, CPA/TCPA, and uncertainty), environmental context (traffic density, visibility, weather, and communication quality), and evidence of operator state (interaction latency, workload indicators, intervention history, and trust and SA measures).

These inputs feed an estimator of operator state and an explanation manager that selects among overview, rationale, uncertainty, alternative actions, and explanations for emergency takeover. Outputs are delivered through chart-centered HMI and eHMI channels and recorded as evidence for validation and governance. The framework forms a closed loop linking MASS and GNC inputs and environmental conditions with RO actions and system outcomes. This allows explanation policies to adapt when workload, risk, or communication conditions change. Fig. 9 summarizes the proposed architecture.

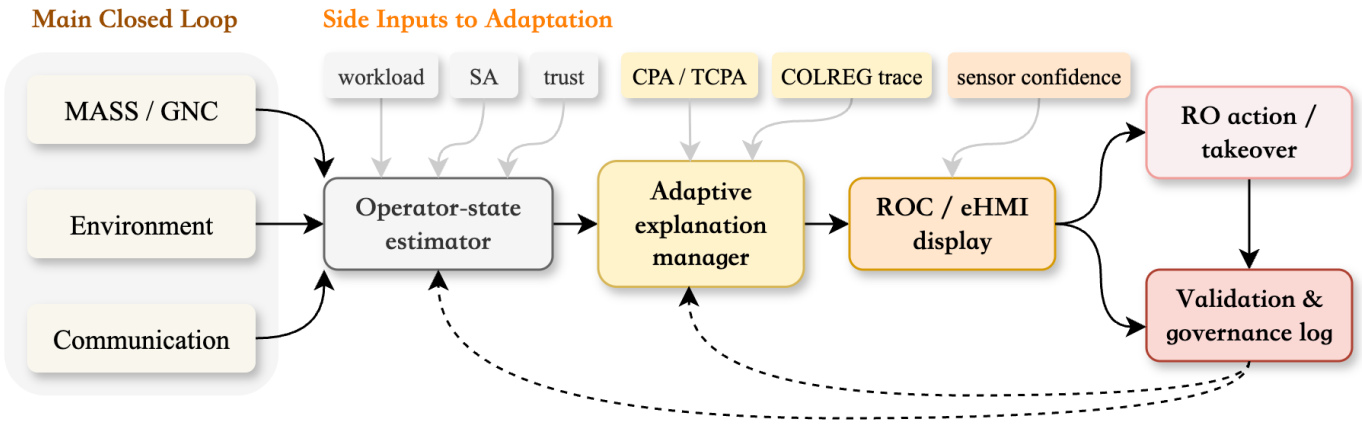

**Fig. 9.** Closed-loop adaptive transparency framework for MASS remote operation.

*2. How to improve transparency for MASS*

a) **SA Data Collection and Visualization**

SA describes how MASS systems acquire, fuse, and interpret environmental information through sensor fusion to create a consistent situational model. Reference [62] combined *failure mode and effects analysis (FMEA)* with risk matrices to identify navigational RIFs in MASS. The framework identified 120 RIFs. Of 17 human-related RIFs in the *Facilities* domain, 12 were high risk. *"Poor situational awareness"* was among them and was also rated high risk in the *Control, Power,* and *Cargo* subsystems. Reference [63] proposed a quantitative SA model in 2018 and found that ROs were significantly more likely than conventional ship operators to lose adequate SA. Technology must therefore account for human limitations and strengths and explicitly support SA [27].

i) *Information collection*

Efforts to improve MASS SA first target perception and sensing. Reference [64] introduced a multimodal neural network for perception that provides high-level SA for collision avoidance. Reference [65] used ceiling cameras and synchronized video workstations to generate vessel transit paths and improve operator SA. To address constraints on edge devices, lightweight YOLO ship detection frameworks also improved the detection of small nearshore vessels by USVs in complex maritime scenes [66].

Ontology frameworks also support SA. Reference [67] integrated information from individual ships, human operators, and a new distributed SA layer into a unified model. The model generates recommendations based on rules and flow control that align with COLREGs. Reference [68] then combined data from multiple sensors and maritime regulations in a knowledge map model with the Dynamic Window Approach for path planning, enabling proactive rather than purely defensive navigation.

State modeling and recognition provide another direction. Reference [69] developed SubCMedians, an unsupervised subspace clustering method that uses data to model and explain operational states such as "in or out of water" and "manual or autonomous". Unsupervised novelty detection dynamically updates the state model library. Higher silhouette scores than those of k-means and Gaussian Mixture Models (GMM) indicate better clustering and interpretability, supporting MASS SA in unsupervised environments.

To improve sensing coverage and robustness, [70] proposed cooperative sonar detection using multiple USVs, which covered a wider area than a single USV. By analyzing temperature, sound speed gradients, and propagation loss, and optimizing sonar depth and formation, the study validated dynamic sensor layouts for underwater SA.

Model-based decision frameworks also connect SA with collision avoidance. Reference [71] separated situation interpretation from collision avoidance planning in a framework based on discrete event systems. It couples deterministic finite state automata and synchronizes them across vessels, enabling safe and efficient cooperative decisions among multiple ships while complying with COLREGs.

*ii) Information exhibition*

Information presentation strongly affects transparency and operator trust. Trust falls when decision algorithms are hidden, while transparent text or graphics increase confidence [36]. In ROCs, additional viewpoints such as AR and panoramic cameras can be combined with radar data to give ROs more information and support safer decisions. Reference [72] showed that a simple interface with diagnostic messages and support for prompt responses helps ROs manage operational challenges. Reference [60] also emphasized dynamic situation cognition and navigation intention prediction models for natural interaction between navigators and autonomous systems.

At a design level, [33] proposed a multi-level HMI design framework including informational, communicative, and physical levels. Based on these findings, two main directions for improving information exhibition in MASS have been identified:

***Direction 1: Conceptual eHMI***

*A) Ship-mounted displays*

LED/LCD screens built into the vessel can communicate text and symbolic cues directly to nearby ships [33]. Reference [73] displayed robustness in path following and collision avoidance as live progress bars or heatmaps on navigation charts, allowing operators to assess safety margins under current noise conditions. Computer vision provides another approach by outlining detected obstacles with bounding boxes and mapping them virtually to support operator decisions [74]. Reference [75] designed and simulated a decision support system (DSS) for cooperative collision avoidance between humans and machines. The system quantified potential collision losses and displayed them using different colors to distinguish costs, thereby improving judgement and decision efficiency.

*B) LED Light Strips and Color Coding*

Color coding provides an intuitive way to improve decision transparency. ARPA and ECDIS overlays can display Time to Closest Point of Approach (TCPA) and Closest Point of Approach (CPA) risks using five colors ranging from no hazard to extreme danger [74]. A rosette of colored sectors can similarly indicate maneuverability zones. External LED strips can communicate operational states such as "ready", "warning", and "propulsion initiated" through changing colors and flashing patterns [33]. Combined with fuzzy logic models based on COLREGs and executable algorithms, these displays can present rule compliance scores, probability distributions, and ranked decision alternatives with explanations, such as "Decision based on rule XX, compliance score YY" [76], [77]. Sparse state models can further limit the display to relevant sensor subspaces and provide graphical thresholds and confidence indicators, supporting efficient and transparent HMI design [69].

***Direction 2: Conversational HMIs***

*A) Conversational UI*

Conversational interfaces allow operators to query MASS states and operational intentions through natural language by voice or text [33]. Reference [78] developed a *voice guided situational prompt system* for beginner navigators. Simulator experiments using the *Situational Awareness Global Assessment Technique (SAGAT)* showed that it improved SA and decision making among novice operators. Reference [29] further highlighted that *chart centered interac*tive panel*s* can improve performance by providing access to vessel charts, sensor data, and system health status. In simulations, context specific warnings and alerts displayed on the main sea chart improved operator responsiveness and decision confidence.

*B) Immersive Visualization (VDES, VR, AR)*

Immersive visualization using VDES remote communications, VR, and AR may offset the loss of "ship sense" among ROs. Ship sense is the tacit awareness normally supported by auditory, vestibular, and tactile cues [56]. Workshops and simulator studies, including ARSVO [79], show that missing sensory feedback can make remote control feel like a game, causing psychological detachment and reduced situational realism [36]. Proposed responses include forthcoming VDES e-Navigation capabilities and VR and AR interfaces that directly visualize MASS operations [33]. Survey results also revealed demand for these approaches [80]. Eleven of 32 participants endorsed onboard cameras with a 360° view to avoid frequent changes in camera angle, while five requested large displays showing all vessels simultaneously. Both were expected to improve SA and reduce cognitive load.

Practical MASS HMI cases include NTNU's AutoFerry sketches [81], which combine an egocentric 360° camera view with an exocentric map integrating radar, LiDAR, and AIS. A program panel displays predefined "Routine" and "Emergency" automation steps, highlights active tasks, and allows manual override and fine adjustment with a joystick. Reference [53] proposed the Quickly Getting Into the Loop Display (QGILD), a standardized format that uses colored alerts and concise intervention options to present deviations. Maps or camera views further support rapid recognition and recovery based on operator skill under stress.

To address limitations of frameworks such as KONECT, [82] developed an extended interaction process that uses Keystroke Level Model (KLM) efficiency to optimize UI components and help designers select effective elements for specific tasks. For external HMIs, [83] introduced the ZWIPP eHMI, which uses 24 rotating panels and LED strips to display 13 dynamic status and intent messages. It retracts to minimize its visual profile and deploys to communicate with nearby vessels or shore personnel. Validation, however, remains limited to 3D simulations and excludes real-world and auditory interaction testing.

New methods include graph theory modeling of human–machine RIFs [61], interpretable collision avoidance systems designed for HMI [84], and GPU accelerated AIS trajectory compression for visualization with low latency [85]. However, many solutions depend on weather, lack regulatory frameworks, and remain at the conceptual or simulation stage [33]. Evaluations also show that transparency layers do not consistently improve SA. Designers must therefore balance transparency against cognitive workload to support effective decisions without overwhelming operators [86].

b) **Engineering design strategies**

*i) Resilient interaction design (resilient IxD)*

Reference [87] introduced resilient IxD to address challenges in interactions between humans and AI within complex systems such as transportation. In this context, transparency depends on both the ROC interface and the constraints that shape engineers' behavior during system development. Engineering organizations can support this work by improving informal information sharing through dedicated gatekeepers or organizational liaisons [29].

Validation is central to design focused on engineers. The Think Aloud Protocol used in [29] reveals operator reasoning in "unknown unknown" scenarios and provides feedback to designers and engineers. Reference [88] proposed an evaluation across multiple phases that adapts to equipment degradation and changes in system composition, ensuring that decisions reflect the vessel's actual health. The Risk Appetite based Collision Avoidance Decision Making System (RA-CADMS) in [89] combines human risk tolerance with computational optimization. By incorporating crew risk preferences while complying with COLREGs, it produces more interpretable avoidance decisions aligned with operators and represents an early attempt to integrate risk attitudes into collision avoidance automation. Future work should add factors such as sea state and vessel type and validate the model through trials using full-scale vessels to improve the fidelity and transparency of decision data.

*i) Predicting HMI Operational Errors*

Decision transparency requires the decision itself to be clear, especially in collision avoidance, where trajectories, routes, and speeds are often adjusted [74]. Accurate prediction of HMI operational errors is therefore essential to MASS transparency and operator reliability.

Reference [80] found that improved tracking based on Closest Point of Approach (CPA) was the most common DSS recommendation for enabling early collision warnings, reported by 17 of 32 participants. Operator errors have also been quantified through several models. Reference [90] used the Success Likelihood Index Method (SLIM) with interval type 2 fuzzy sets (IT2FSs) to aggregate expert estimates of human error probability (HEP), enabling robust and flexible prediction under uncertainty. Reference [91] complemented this approach with catastrophe theory, modeling nonlinear risk dynamics to identify tipping points at which operational states may deteriorate into accidents. For MASS research with humans in the loop (HITL), [92] integrated compact Kongsberg simulators, full mission bridge simulators, immersive simulators, and the R/V Gunnerus through a unified data exchange interface. Its five stages covered design, verification, deployment at scale, data analysis, and validation. Multimodal EEG and eye tracking data, together with machine learning models including SVM and random forest, captured cognitive load and operational behavior. This created a closed loop from expert experience to decision support based on data. Differences in simulator interface standards nevertheless limited generalizability, showing the need for standards across platforms to improve scalability and practical adoption [93].

Reference [94] compared two kinematic models, CMM and CTRA, and two Kalman filters, EKF and UKF, using Monte Carlo simulation, $\chi^2$ tests, and autocorrelation tests. CTRA produced smoother and more robust trajectory predictions during abrupt maneuvers, while EKF achieved accuracy comparable to UKF with greater computational efficiency. These results show that advanced trajectory modeling and uncertainty visualization can improve MASS transparency through smoother state predictions, clear confidence indicators, and adaptable configuration options that do not overload operators.

*ii) Vessel Emergency Plan (VEP)*

Displaying recommended routes or maneuvers on navigation maps helps operators understand system intentions [74]. These visualizations should also explain how decisions were produced while preserving SA and avoiding information overload [27]. For safe operation without onboard personnel, MASS also require a dedicated *Vessel Emergency Plan (VEP)* that can be activated and executed from the ROC. The VEP should show the planned response and explain contingency strategies for both routine and emergency scenarios [50].

*iii) e-Navigation simulation experimental systems*

MASS development requires models that integrate data with system mechanisms. Unlike autonomous vehicles, which benefit from public datasets, navigation research on human–machine cooperation lacks a comprehensive database and standardized platform. To balance cost and risk, [60] recommended a navigation simulation system incorporating *Hardware* i*n the Loop (HIL)* and *Man* i*n the Loop (MIL)*. Supporting frameworks include a *discrete event system (DES)* that separates SA from decision making when predicting the behavior of multiple ships [95], and an *integrated 3D digital twin* for obstacle detection and path optimization [96]. Connectivity remains difficult in remote regions such as the Arctic, where AIS heartbeat and telemetry data are often discontinuous. Reference [97] proposed a hybrid *direct to satellite (DtS)* LPWAN/mMTC architecture using LoRaWAN or LR-FHSS to transmit AIS heartbeat data directly to low Earth orbit (LEO) satellites. This architecture significantly expands MASS monitoring coverage and provides more reliable, continuous SA, thereby supporting transparency in extreme environments.

The reviewed solutions vary in maturity. Conceptual HMI/eHMI proposals and ontology-based decision models clarify design requirements, but many remain prototypes or 3D simulations. Simulator studies provide stronger evidence on workload, SA, and takeover behavior. MIL, HIL, HITL, and digital-twin validation also provide evidence on interface timing, sensor degradation, and data interfaces before field trials. Evidence from operational deployment is still limited. Future safety cases should therefore distinguish conceptual design, simulator evaluation, HIL/MIL/HITL validation, field trials, and operational monitoring [57], [92], [93], [104].

**c) Remote Operator Behavior and Training**

ROs must avoid excessive reliance on automation and retain the ability to verify system performance independently. Understanding the underlying technology is essential for detecting anomalies and maintaining safe operations [98]. Navigational competence is therefore fundamental to future RO training models, with *analytical hierarchy process (AHP)* studies ranking it as highly important [18]. Training should also present predefined scenarios transparently and explain how MASS systems make decisions. This improves information transparency in maritime training and strengthens MASS explainability.

For certification, RO competence should be demonstrated through evidence rather than course completion alone. Core competencies include remote supervision, anomaly diagnosis, takeover and emergency control, ship-shore communication, COLREG interpretation, cyber/data awareness, and coordination with onboard or nearby human actors. Scenario-based simulators should be combined with HIL/HITL evidence under degraded sensors, delayed communications, multi-vessel supervision, and emergency handover conditions, building on STCW competence logic while adapting it to ROC work [57], [92], [93], [98], [107].

Risk modeling for MASS is constrained by the lack of historical data for this emerging field. *STPA* and *BNs* partly address this limitation, but empirical validation remains necessary [42]. Training through simulation offers one solution. Immersive virtual environments can evaluate interactions between humans and machines and among humans, improve operators' situational understanding, and test interaction strategies before deployment. The main virtual training methods are:

*i) Early-stage scripted scenarios training*

Scripted virtual simulations improve operators' cognitive performance in hazardous environments by exposing them to predefined abnormal situations. These scenarios exercise the diagnosis and action execution functions in the IDAC model and help operators form accurate mental models of system dynamics and intervention timing. The approach has been applied in rail, medicine, and automated vehicles [29], [36]. Maritime studies also indicate benefits for RO fault recognition and SA [72]. However, the validity of these cognitive models remains contested because human decision making is complex and variable. Neuroscience based approaches may therefore be needed to strengthen maritime risk modeling [42].

*ii) Generative AI-Based Training*

Generative AI training prepares ROs to use advanced MASS technologies, including AI systems, IoT, VR/AR, and data analytics. It requires social skills such as communication and cognitive skills such as understanding AI principles. *Large Language Models (LLMs)*, including *SeaGPT*, demonstrate the need for skills in communication automation through interactions between crew managers and port agents [25]. To support cooperative decisions involving humans and machines, [60] proposed a self learning framework based on parallel systems and *BiLSTM neural networks*.

However, simulator training can foster "*implicit trust*" in automation. Reference [79] found that although 82.5% of cadets and 66.6% of experienced navigators reported only slight trust, many continued to trust compromised systems during initial simulations. This bias was partly caused by simulator configurations prepared by instructors, which trainees rarely questioned. Survey results also showed that 20% of participants were neutral and 5% disagreed that training improved trust calibration. Training may therefore fail to change operator behavior and may reinforce existing biases. This highlights the need for training focused on critical thinking and frequent emergency response tests.

**d) Regulatory and Rule-Based Improvements**

*i) Standardizing Ship-to-Ship Communication*

Clear communication of MASS intentions to surrounding vessels is essential for safe navigation. Route exchange allows ships to share planned routes and intended deviations through VHF Data Exchange Systems (VDES), so that both humans and machines can incorporate this information into their decisions [74].

Standardization is needed because conventional and remotely operated ships work differently. Conventional crews rapidly coordinate avoidance maneuvers by radio, flares, or sirens. ROs often decide without direct contact with nearby vessels, increasing confusion and misunderstanding [50].

To address this gap, [99] proposed an ontology-based decision algorithm for communication between autonomous ships and between autonomous and manned vessels. Simulations showed that the algorithm produced unambiguous solutions when Collision Regulations could be interpreted differently. This increased navigators' trust in autonomous vessels and supported maritime safety.

*ii) AI-powered collision avoidance systems and COLREG compliance*

ROs may understand the "rules of the road", but trusting an autonomous ferry to follow them is a separate challenge [100]. Fault Tree Analysis of 513 reported collisions identified COLREG violations as the largest contributing factor [101]. COLREGs were designed for human involvement, and many provisions remain linguistically ambiguous. Terms such as "safe distance" and "shall give way" lack standard quantitative definitions, which complicates formalization and algorithmic implementation [102]. Verifying that autonomous algorithms interpret and apply COLREGs reliably therefore remains a major safety challenge.

Reference [76] used *fuzzy logic* to quantify linguistic ambiguity in COLREGs. Qualitative rules, including give-way obligations and safe passing distances, were mapped to fuzzy membership functions from 0 to 1. Variables such as relative distance, relative speed, and maneuvering capability then produced an interpretable "*avoidance degree*", such as "strongly avoid" or "slightly adjust", with confidence levels. Reference [77] combined fuzzy logic with a *risk graph model* for scenarios involving multiple ships, conflicts, and criteria. The executable algorithm visualizes decisions so that ROs can understand and trust AI strategies for collision avoidance in complex environments. Reference [103] addressed changing encounter risks with a rule-aware time-varying conflict risk measure. It combines COLREGs with *maritime practice* and quantifies risk changes in line with RO expectations. Compliance with COLREGs, however, is only part of regulation. Reference [104] examined *safety acceptance principles* for autonomous ships. Regulators also need clear, measurable safety thresholds to certify that AI-driven systems are *safe* for deployment [105].

Comparison with other domains separates shared challenges from those specific to MASS. Like autonomous vehicles and aviation, MASS faces mode confusion, takeover pressure, and trust calibration problems when humans supervise highly automated systems [26], [36], [105]. MASS differs because an RO may supervise several vessels through delayed or degraded communications, while COLREG obligations in mixed traffic remain ambiguous and depend on context. Recent USV surveys also reveal fragmented methods and applications across platforms, highlighting the need to link decision transparency with operational validation and governance evidence [142].

## V. Conclusion

This review of 100 studies examines how transparency and explainability improve MASS safety and usability. It links the GNC stack to shore operations and identifies Human-UCA concentrations during transitions from RSM to RCM and emergency loops. Evidence from human factors, HMI/eHMI design, and regulation shows that decision rationales, variables and weights, uncertainty and confidence, and ranked executable alternatives improve understanding and trust calibration. Trust also depends on reliability, predictability, and task support.

The framework addresses validation and communication. Validation requires common, validated trust and SA instruments, behavioral and performance indicators, workload proxies, and preregistered mixed effects analyses across simulation, HIL/MIL/HITL, and digital-twin testbeds. Communication requires XAI aligned with user mental models and delivered through chart-centered or external HMIs. Examples include CPA/TCPA risk widgets, robustness indicators, COLREG rule traces with confidence, and VDES intent sharing. The framework also adapts explanation detail and timing to operator state, using behavioral, performance, and physiological signals to prevent overload or insufficient information.

Three near-term priorities support practice and policy: deploying explainable decision-support displays in existing bridge and ROC workflows, adding transparency requirements and evidence to safety cases and the emerging MASS Code, and advancing COLREG formalization and standardized route and intent sharing to reduce ambiguity in mixed traffic. COLREG methods may include fuzzy and risk-graph pipelines.

The study has three limitations. Coverage ends in early 2025, the corpus is heterogeneous, and operational data remain scarce. Future work should develop shared scenario libraries and benchmarks, open audit trails linking explanations to operator responses and outcomes, and multi-site trials measuring takeover timeliness and Human-UCA rates. Empirically validated transparency, clearly communicated through adaptive HMI, offers a practical path to trustworthy human–machine collaboration in MASS.

## APPENDIX I

## THE SELECTED STUDIES

| *Reference* | *Method* | *Topic* |
|---|---|---|
| Ramos et al., 2019 [24] | Qual.: Hierarchical Task Analysis (HTA), IDAC | Collision avoidance |
| Alsos et al., 2022 [33] | Qual.: SME Interview study | HMI, Automation transparency |
| Asplund & Ulfvengren, 2022 [29] | Qual.: Think-aloud Protocol (TAPI) Interviews | Human-centered design, Emergency response system |
| Banda et al., 2021 [106] | Qual.: Framework development | Risk management |
| Cheng et al., 2023 [38] | Qual.: System Theoretic Process Analysis (STPA-Cog), IDAC | Human-system collaboration, Hazard identification |
| Danilin et al., 2022 [135] | Qual.: Case Incident Analysis | Cyber security |
| Emad & Ghosh, 2023 [98] | Qual.: In-depth Interviews | MASS operator training |
| Koskinen et al., 2024 [111] | Qual.: Participatory Design | Bridge zero, Concepts of operations |
| Song et al., 2023 [67] | Qual.: Ontology-based modelling | Situational awareness |
| Goerlandt, 2020 [140] | Qual.: IRGC Risk Governance Framework application | Decision support system, Safety, Risk assessment |
| Heikkilä et al., 2017 [138] | Qual.: Goal-based Safety Case Approach | Safety, Situational awareness, Potential hazards |
| Lutzhoft et al., 2019 [116] | Qual.: Expert workshops | Cyber security, Human factors, HMI |
| Lynch et al., 2022b [36] | Qual.: Perceptual Cycle Model (PCM) | Situational awareness, Decision making |
| Hamrén et al., 2024 [137] | Qual.: Edge AI prototype development & demonstration | Situational awareness |
| Man et al., 2021 [118] | Qual.: Focus group interview study | Human factors. Situational awareness |
| K. Lynch et al., 2023 [59] | Qual.: Schema World Action Research Method (SWARM) | Decision making, Trust |
| Porathe, 2022 [53] | Qual.: Concept prototype design & expert walkthrough | Human-in-the-loop, HMI |
| Petermann et al., 2024 [122] | Qual.: Participatory design | Hazard management, Human-centered design |
| Johnsen et al., 2022 [109] | Qual.: Hazard Identification & Risk Assessment | Risk assessment, Safety, Human factors |
| Porathe, 2021b [81] | Qual.: Concept sketch | Human factors, HMI |
| Sunkara et al., 2023 [124] | Qual.: Simulation-based experiment for two cases | Design tradeoffs, Situational awareness |
| Van Den Broek & Van Der Waa, 2022 [100] | Qual.: Dynamic task allocation algorithm | Risk assessment, Situational awareness |
| E. Veitch et al., 2020 [128] | Qual.: Interview study | Human-centered design |
| Veitch et al., 2021 [87] | Qual.: HCD design | Human-system interaction in autonomy (H-SIA) |
| Wu et al., 2024 [93] | Qual.: Simulation-based experiment | Human-in-the-loop |
| Wu et al., 2022 [92] | Qual.: Simulation-based experiment | Human-in-the-loop, Monitoring, HMI |

| *Reference* | *Method* | *Topic* |
|---|---|---|
| Bakdi & Vanem, 2022 [76] | Quant.: Linguistic Fuzzy Inference System | Collision avoidance, HMI, COLREGs |
| Bakdi & Vanem, 2024 [77] | Quant.: Graph model for conflict resolution | Collision avoidance, COLREGs |
| S. Fan & Yang, 2023 [61] | Quant.: Tree Augmented Network (TAN) modelling | Human factors, Risk influencing factors |
| Feng et al., 2023 [70] | Quant.: Multi-sonar collaborative detection algorithm | Situational awareness |
| Garofano et al., 2022 [64] | Quant.: MILP-based path planning; computer-vision perception; PID & MPC control | Collision avoidance, Mixed Linear-Integer Programming, Model-based design |
| Hansen et al., 2020 [71] | Quant.: Discrete-Event Systems Modeling | Situational awareness, COLREGs |
| Gash et al., 2020 [65] | Quant.: Vision-based obstacle detection & path planning algorithm | Situational awareness, Path planning |
| Hinostroza & Lekkas, 2024 [139] | Quant.: Deep learning–based risk prediction model | Temporal AI planning, Intelligent navigation |
| Castellini et al., 2019 [69] | Quant.: Subspace Clustering (SubCMedians) | Situational awareness, Sensor data |
| Chan et al., 2022 [72] | Quant.: Simulation-based experiment, Event tree analysis | Situational awareness, Human factors |
| Lee et al., 2023 [113] | Quant.: Fault Tree Analysis | Collision Avoidance, Risk analysis |
| Lee et al., 2024 [73] | Quant.: Deep reinforcement learning with digital-twin simulation & Gaussian noise modeling | Collision Avoidance, Decision making, Safety, Robustness |
| J. Liu et al., 2021 [90] | Quant.: Success Likelihood Index Method (SLIM-IT2FSs) | HMI, Human factors |
| Y. Li et al., 2024 [85] | Quant.: Adaptive DP with Speed and Course (ADPSC) algorithm & GPU parallel computing framework | Trajectory compression, Safety, Parallel computing |
| Mei et al., 2023 [75] | Quant.: DSS based on Decision Disc (DD), FSM & RVO-based algorithm | Collision avoidance, Path planning, HMI |
| Man et al., 2018 [56] | Quant.: Controlled drilling experiment | Human factors. Situational awareness |
| Menges et al., 2023 [119] | Quant.: Digital Twin Modeling & Simulation Evaluation | Situational awareness, Digital twin |
| Papageorgiou et al., 2022 [95] | Quant.: Nested Finite Automata modelling | Situational awareness, Finite-state automata |
| Perera, 2018 [120] | Quant.: Deep learning–based navigation model training | Navigation system, Decision making |
| Perera, 2019 [121] | Quant.: Deep learning model evaluation | COLREGs, Situational awareness |
| Pietrzykowski et al., 2024 [99] | Quant.: Ontology-based decision-making algorithm | Human-machine communication, Navigation |
| Kim, 2024 [18] | Quant.: Survey & Analytic Hierarchy Process | Remote operator |
| Song et al., 2024 [68] | Quant.: Ontology-based knowledge maps with the Dynamic Window Approach (KM-DWA) | Situational awareness, Decision making, Knowledge map, Collision avoidance |
| Raza et al., 2022 [96] | Quant.: DRL algorithm | Collision avoidance, Digital twin, Situational awareness |
| Wang et al., 2023 [62] | Quant.: Failure mode and effects analysis (FMEA), Hierarchical model | Risk influencing factor |
| Zhou et al., 2019 [63] | Quant.: Probabilistic theory, SA Quantitative model | Situational awareness, Navigation safety |

| *Reference* | *Method* | *Topic* |
|---|---|---|
| Y. Wang et al., 2023 [94] | Quant.: Kinematic motion models (CMM/CTRA), Monte-Carlo based simulation | System state estimation, Continuous-discrete models |
| Adnan et al., 2024 [50] | Exp: Simulation-based Experiment | Situational awareness, ROC |
| Gregor et al., 2023 [141] | Mixed: VR immersion experiment with SAGAT & trust surveys | Trust, Situational awareness, VR |
| Huang et al., 2020 [84] | Mixed: HMI-CAS framework | Collision avoidance, HMI |
| S. Fan et al., 2024 [30] | Mixed: Graph Theory-based Network Analysis of RIFs | Safety, Human factors, Human-machine system |
| P. Li et al., 2024 [47] | Mixed: Fault Tree Analysis & Fuzzy Bayesian Network (FTA-FBN) | Collision Avoidance, Bayesian network, Fault tree analysis, Fuzzy theory |
| Lamm et al., 2022 [112] | Mixed: Architecture design & teleoperation performance experiments | Situational Awareness |
| A. Madsen et al., 2023 [35] | Mixed: Simulation-based experiment with ECDIS & AIS, Survey study | Decision transparency, XAI, Human, Machine collaboration |
| Jp et al., 2022 [52] | Mixed: Full-bridge simulator experiments & Survey study | Cyber security, Trust, Situational awareness |
| Nishizaki & Hamanaka, 2024 [78] | Mixed: Voice-guided situational prompts & decision logging | Situational awareness, Decision making |
| A. N. Madsen et al., 2025 [86] | Mixed: Simulation-based experiment with SAGAT, User satisfaction survey study | HMI, Interaction design |
| Misas et al., 2024 [79] | Mixed: Full bridge simulator exercises & Interview study | Trust, Cyber security, Situational awareness |
| Troupiotis-Kapeliaris et al., 2023 [126] | Mixed: Neural network & Simulation-based experiment | Design, Collision avoidance |
| Ullah et al., 2022 [97] | Mixed: mMTC DtS architecture & Simulation by LoRAN | Situational awareness, Communication system |
| Vagale et al., 2022 [57] | Mixed: Simulation-based Experiment with human-in-the-loop | Human-in-the-loop, Maritime training |
| Saager et al., 2024 [82] | Mixed: Extension of KONECT method & User evaluation study | Human factors, Interaction-designed, Transparency |
| E. Veitch et al., 2024 [80] | Mixed: Factorial simulation-based experiment, ANOVA, Interview | Human factors |
| Simic & Alsos, 2023 [83] | Mixed: 3D CAD prototyping with KeyShot rendering & Survey | Automation transparency, External HMI |
| X. Wu et al., 2021 [89] | Mixed: RA-CADMS Framework, Decision matrix, Simulation | Collision avoidance, Decision making, HMI |
| Han et al., 2021 [44] | Mixed: MADS-IDAC system, Simulation-based experiment | HMI, HCI design, Knowledge-based systems |
| R. Zhang & Furusho, 2020 [134] | Mixed: AIS ship navigation environment, Deep reinforcement learning | Risk Perception, Collision avoidance, Situational awareness |
| Zhang et al., 2022 [132] | Mixed: Hybrid causal logic (HCL), Fault tree (FT), Bayesian Belief Network (BBN) | Risk assessment, Hazard identification, Human factors |
| S. Li & Fung, 2019 [114] | Literature Review | Implementation, Safety |
| Y. Li et al., 2019 [115] | Literature Review | Risk assessment, Operator competence |

| *Reference* | *Method* | *Topic* |
|---|---|---|
| Guevara & Dalaklis, 2021 [136] | Literature Review | Safety, Risk assessment |
| Kokubun, 2022 [110] | Literature Review | Safety evaluation |
| Kristoffersen, 2020 [27] | Literature Review | Human-centered design, Human error |
| MacKinnon et al., 2020 [117] | Literature Review | Decision making, Human factors |
| Fan et al., 2020 [45] | Literature Review | Safety, Risk influencing factor |
| Forti et al., 2022 [22] | Literature Review | Automatic Anomaly Detection, Risk assessment |
| Aslam et al., 2020 [39] | Literature Review | Situational awareness |
| Chaal et al., 2023 [28] | Literature Review | Safety, Reliability, Risk assessment |
| Belabyad et al., 2025 [25] | Literature Review | Human factor, Maritime training |
| Madsen & Kim, 2024 [74] | Literature Review | Transparency, Decision making |
| Liu et al., 2022 [60] | Literature Review | HMI, Situational awareness, Decision making |
| Lynch et al., 2022 [36] | Literature Review | HMI, Transparency, Situational awareness |
| Song et al., 2024b [32] | Literature Review | Trust, HMI, Situational awareness |
| Porathe, 2021 [34] | Literature Review | Human factors, HMI |
| Tao et al., 2024 [125] | Literature Review | Safety, Risk analysis |
| Alamoush & Ölçer, 2025 [8] | Literature Review | Collision avoidance, Situational awareness |
| Ranjan et al., 2025 [31] | Literature Review | XAI, Trustworthiness |
| X. Li & Yuen, 2024 [7] | Literature Review | Human factors |
| E. Veitch & Alsos, 2022 [42] | Literature Review | HCI, Safety, Resilience engineering, Interaction design |
| Wahlström et al., 2015 [130] | Literature Review | Human factors, Monitoring |
| J. Zhang et al., 2023 [133] | Literature Review | Machine-type communication |
| Wróbel et al., 2022 [102] | Literature Review | Collision Avoidance Methods (CAMs), COLREGs, Safety |
| Wróbel et al., 2021 [131] | Literature Review | Collision avoidance, Intact stability, Communication |
| Saghafian et al., 2024 [123] | Literature Review | Automation transparency, Situational awareness, Trust |

## Acknowledgment

The author would like to thank Prof. Haitong Xu for valuable discussions on the review process and system architecture, and Prof. C. Guedes Soares for his careful feedback. The author also thanks the Centre for Marine Technology and Ocean Engineering (CENTEC) for providing a supportive academic environment. The author is grateful to the anonymous reviewers for their constructive comments, which helped improve the quality of this article.

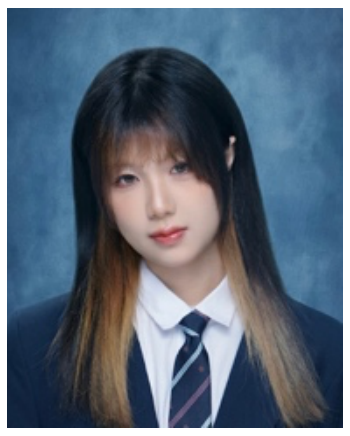

**Zhuoyue Zhang** received the B.S. degree in the department of Applied Mathematics in the Chinese University of Hong Kong, Shenzhen, and the M.S. degree in department of Civil Engineering in KTH Royal Institute of Technology and Instituto Superior Técnico.

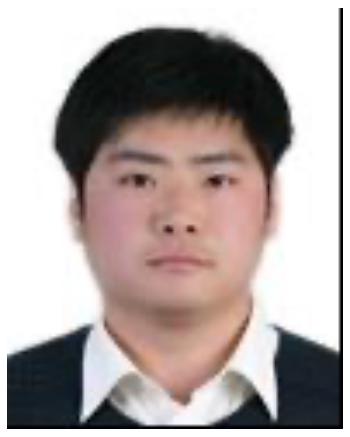

**Haitong Xu** is an Assistant Professor of Instituto Superior Técnico, University of Lisbon and he received the Ph.D. degree in Naval Architecture and Ocean Engineering, Instituto Superior Técnico, University of Lisbon, Portugal, in 2021. He currently oversees the research on autonomous ships at Centre for Marine Technology and Ocean Engineering (CENTEC), focusing on guidance navigation and control (GNC), path following control, mathematical modeling, data acquisition (DAQ) and developing autonomous ship prototypes.

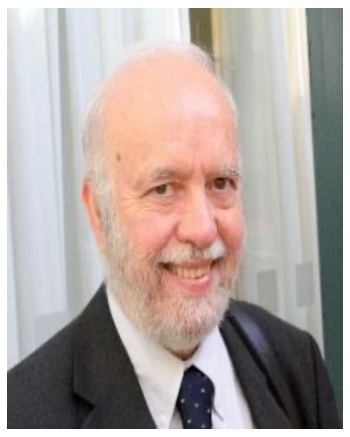

**Carlos Guedes Soares** is a Distinguished Professor of the Engineering School (Instituto Superior Técnico) of the University of Lisbon. He concluded his postgraduate studies at the Massachusetts Institute of Technology, USA, in 1976 and at the Norwegian Institute of Technology of the University of Trondheim in 1984. Since then, he has been at the University of Lisbon (Technical University of Lisbon until 2013). Among his research areas, he has been interested in maritime safety, modeling of maritime traffic and autonomous ships.